\documentclass[11pt]{article}
\usepackage[margin=1in]{geometry}
\usepackage[T1]{fontenc}
\usepackage[utf8]{inputenc}
\usepackage{lmodern}
\usepackage{microtype}
\usepackage{csquotes}
\usepackage[hidelinks]{hyperref}
\usepackage{xcolor}
\usepackage[hidelinks]{hyperref}
\usepackage{graphicx}
\usepackage{float}
\usepackage{subcaption}
\usepackage{xurl}

\usepackage{authblk}

\usepackage[backend=biber,style=authoryear-comp,sorting=nyt,maxcitenames=2,maxbibnames=99,uniquename=false,doi=true,isbn=false,url=true,eprint=false,giveninits=true]{biblatex}
\title{The directions of science: How we search the unknown}
\author{Nicola Cinardi\thanks{Corresponding author:
\texttt{cinardinicola@gmail.com}}\\\small Center for Computational and Stochastic Mathematics, \small
Instituto Superior T\'ecnico, Lisbon, Portugal}
\date{}
\begin{document}
\maketitle
\begin{abstract}
Science is producing knowledge at unprecedented scale, while artificial intelligence promises to accelerate the generation of hypotheses, analyses, and scientific outputs further. Yet increasing the volume of research does not necessarily expand the range of directions science effectively explores. We argue that science can be understood as a collective search process in which finite resources are allocated across an evolving space of possible research trajectories. Careers, funding, organization, publication, attention, and technology act as interacting selection mechanisms that shape this allocation. This Perspective distinguishes the visible diversity of scientific activity from its effective diversity: apparently different researchers, institutions, or projects need not provide genuinely different routes of exploration or evaluation. Our principal contribution is to introduce \textit{cross-stage dependence}, which occurs when selection processes operating at different stages of science are not fully independent in the grounds through which they favour scientific trajectories. We hypothesize that, under uncertainty, repeated filtering through overlapping criteria, information, prior selections, or technological infrastructure can reduce the capacity of science to maintain plausible alternatives and redirect effort when evidence changes their relative promise. Artificial intelligence makes this problem increasingly consequential because it can simultaneously expand scientific search capacity and participate across multiple stages of production and selection. The challenge is therefore not simply to produce more knowledge, but to preserve sufficiently distinct routes of exploration, verification and correction while retaining the ability to change direction.
\end{abstract}


\section{The paradox of abundance}
Science has become extraordinarily productive. The number of researchers, publications, collaborations, datasets and computational tools has expanded across much of the scientific system. Artificial intelligence may accelerate this expansion further by assisting literature discovery, hypothesis generation, modelling, experimentation, analysis and scientific writing.

It is tempting to interpret this growth as a straightforward expansion of scientific progress. But output and progress are not equivalent. Recent work has reported a decline in the disruptive character of papers and patents, although the interpretation and measurement of disruption remain contested \parencite{park2023papers}. Large scientific fields can also grow while concentrating attention around established work, slowing the incorporation of new ideas into the canon \parencite{chu2021slowed}. More generally, an increase in the amount of science tells us little by itself about the distribution of effort across possible directions.

Consider two scientific systems that double their research output. In one, the additional capacity opens previously neglected questions, methods, and combinations of ideas. In the other, it intensifies activity around directions already attracting most researchers, funding, and attention. Both systems become more productive. Only the first necessarily broadens the search. This distinction becomes more important as scientific output grows because some resources do not expand proportionally. Researchers still have finite attention. Funding remains limited. Journals, institutions and scientific communities must decide which results to evaluate, communicate and retain. As the production of candidate knowledge becomes cheaper, the allocation of attention and judgment can become more consequential rather than less.

By a scientific direction, we mean a trajectory of research defined broadly enough to include combinations of questions, hypotheses, methods, representations, systems and empirical domains. Directions are not fixed objects. New knowledge, technologies and discoveries continually create possibilities and make others obsolete, infeasible or uninteresting. The question is not whether science produces enough novelty, takes enough risk, or maximizes diversity. Mature theories can deserve concentrated effort, reliable methods can appropriately dominate inferior alternatives, and strong evidence should cause scientific communities to converge. The deeper problem is how science distributes finite search capacity before the value of competing possibilities is known.

This problem has important precedents. The division of cognitive labour has long been studied as a problem of how scientific communities distribute effort across competing approaches \parencite{kitcher1990division}. Models of epistemic landscapes have examined how individual and social search strategies influence collective discovery \parencite{weisberg2009epistemic}, while subsequent work has shown that the advantages of diversified search depend on assumptions about landscapes and search behaviour \parencite{alexander2015epistemic}. Empirical research has meanwhile examined how careers, funding, teams, institutions, publication and attention shape scientific activity. Recent systems-level analyses have argued that these mechanisms can interact, with short funding cycles, productivity-based evaluation, career precarity and conservative selection creating incentives for safer and more readily evaluated research \parencite{haubrock2026contemporary}. These literatures have largely examined how particular mechanisms or combinations of institutional pressures shape scientific behaviour. Less attention has been given to whether selectors operating at different stages provide genuinely distinct judgments or repeatedly reproduce overlapping selections.

We therefore shift attention from the number of researchers, projects or institutions to the mechanisms through which scientific trajectories are selected. The central question becomes: How many genuinely distinct routes of exploration and evaluation does a scientific system preserve? A field may contain many funders, laboratories and journals yet provide fewer independent routes of selection than these numbers suggest if later decisions repeatedly inherit signals created by earlier ones. The issue is increasingly important because the same technological systems can now participate in several parts of the scientific process. AI tools can influence what researchers read, which hypotheses they generate, how analyses are performed, how results are written, and potentially, how scientific work is evaluated.
Scientific abundance therefore creates a paradox. The capacity to produce research can increase at the same time that scientific search becomes concentrated around fewer effective routes. Existing work has extensively studied diversity among scientists, cumulative advantage, path dependence, and the effects of individual selection mechanisms. Here we focus on a different property: whether functionally different mechanisms of scientific selection provide genuinely independent grounds for determining which research trajectories continue. Understanding the future of science requires studying not only how much knowledge is produced, but how the mechanisms that select scientific directions are connected.

\section{Science as selective search} 
At any moment, science contains more possible questions, experiments, models, combinations and applications than can be pursued. Scientific activity therefore necessarily involves selection. This selection differs from search over a fixed landscape because the space of possibilities itself evolves. Discoveries generate new questions, instruments make previously inaccessible phenomena observable, new theories reorganize existing problems, and methods developed in one field create possibilities in another. Conversely, abandoned expertise, infrastructure, or data can make some trajectories difficult to reopen.

Scientific search also occurs under substantial uncertainty. Researchers frequently cannot know in advance whether a question is solvable, whether an experiment will work or whether a currently marginal idea will later become central. A division of cognitive labour can help under such conditions. Different researchers pursuing different approaches can prevent a community from committing prematurely to one trajectory \parencite{kitcher1990division,weisberg2009epistemic}. But diversity is not automatically beneficial. Its value depends on the structure of the problem, the reliability of available evidence, and the costs of maintaining alternatives \parencite{alexander2015epistemic}.

Novel advances also frequently arise through unusual combinations of existing knowledge. Atypical combinations can be associated with high-impact work \parencite{uzzi2013atypical}, and researchers positioned outside established intellectual structures can sometimes generate unusually novel combinations \parencite{shi2023surprising}. Scientific progress therefore depends partly on maintaining access to possibilities that are not already strongly favoured by dominant trajectories.

The problem resembles an exploration--exploitation trade-off, with an important complication: scientific trajectories are not equally reversible. A community can reduce activity in a field and later return to it, but the cost depends on whether expertise, infrastructure, datasets, methods and institutional memory have survived. Temporary neglect of one direction may be inexpensive, whereas abandonment of another may eliminate options that are costly to reconstruct.

To analyse how scientific trajectories are selected, we distinguish four functional stages. \textit{Entry} concerns who enters, remains in, and advances through scientific careers and therefore which dispositions, skills, and interests populate the research system. \textit{Allocation} concerns which people, questions, and projects receive scarce resources such as funding, positions, infrastructure, and time. \textit{Production} concerns which possibilities are transformed into experiments, analyses, collaborations, tools, and scientific outputs. \textit{Retention} concerns which results and trajectories persist through publication, attention, citation, replication, incorporation into methods, institutions, and subsequent research.

These stages are functional categories, not mutually exclusive institutions. A university, funding agency, journal, laboratory, or technological platform can affect several stages simultaneously. We use \textit{selector} in the same functional sense: a selector is any mechanism that systematically changes the relative probability that scientifically available possibilities are explored, supported, transmitted, or retained. This definition is not normative. Experimental evidence and institutional prestige can both alter which trajectories survive without being epistemically equivalent. Selection is unavoidable; what matters is how selections interact and whether apparently different stages contribute genuinely distinct grounds for deciding which trajectories continue.

\section{Selectors of scientific direction}
\subsection{Entry}
Scientific direction begins before a project is funded or a paper is written. It is affected by who enters science, which researchers remain, and what kinds of careers are sustainable. Modern research systems depend heavily on temporary scientific labour. Large-scale analyses have documented the rise and importance of transient researchers within scientific careers and production \parencite{milojevic2018changing}. Career structures can therefore influence scientific search by affecting which researchers and research styles survive long enough to develop.
Scientific risk preferences are also shaped socially. \textcite{shibayama2026risk} find that early-career training environments are associated with later orientations toward research risk, including persistent associations between doctoral supervisors' risk-taking and that of their former students. Selection into laboratories and fields can contribute to these patterns, so they should not be interpreted as showing that training mechanically determines scientific behaviour. They nevertheless indicate that scientific dispositions are partly embedded in institutional and social environments. A system can contain thousands of researchers while narrowing its search if career structures systematically favour similar skills, incentives or intellectual strategies.

\subsection{Allocation}
Once researchers and ideas enter the system, resources must be allocated among them. Research grants operate under unusual uncertainty because the value and sometimes even the feasible outcome of a project cannot be specified in advance. Grant design therefore shapes researchers' incentives, application strategies, and subsequent behaviour \parencite{carnehl2025designing}.

Funding decisions can create cumulative advantages. Researchers who narrowly obtain early funding can subsequently become more successful in acquiring research funding than similar researchers who narrowly miss it, illustrating how an allocation decision can influence later opportunities \parencite{bol2018matthew}. Systems that rely heavily on established indicators can also disadvantage work near the intellectual edge. \textcite{packalen2020nih} find that biomedical papers building on very recent ideas are less often NIH funded than papers building on ideas that have had more time to mature, despite a more complicated overall relationship between NIH funding and novelty.

Risk introduces an additional difficulty. Evaluation systems must distinguish poorly designed research from research whose uncertainty follows from genuinely unexplored territory. Evidence from NIH renewal decisions suggests that riskier grants are renewed at lower rates across several measures of scientific risk \parencite{azoulay2025does}. None of this implies that funding should maximize risk or novelty. The issue is whether allocation mechanisms preserve enough plausible alternatives when evidence is insufficient to establish which will ultimately prove valuable.

\subsection{Production}
Resources are transformed into knowledge through scientific organization. One of the most striking long-term changes in science has been the increasing role of teams. Across many fields, team-authored research has grown relative to solo work and is frequently associated with highly cited contributions \parencite{wuchty2007increasing}. Team assembly itself has a systematic structure and can affect creative performance \parencite{guimera2005team}. Part of the broader shift toward collaboration may reflect the increasing burden of knowledge: as fields accumulate expertise and techniques, researchers can respond through greater specialization and teamwork \parencite{jones2009burden}.
But teams are not interchangeable. Large and small teams appear to occupy different positions in the innovation process: smaller teams have been associated more strongly with disruptive work, whereas larger teams more often develop established directions \parencite{wu2019large}. Organizational hierarchy also matters. Flatter teams have been associated with greater novelty and more disruptive research, whereas more hierarchical teams more often develop existing ideas \parencite{xu2022flat}. Large coordinated teams remain indispensable for many forms of contemporary science. The broader point is that organizational structure changes which possibilities can be pursued effectively.
Methods and infrastructure operate similarly. Instruments, datasets, software platforms, and shared analytical procedures expand scientific capacity while making some questions easier to ask than others. Production technologies therefore do not merely accelerate a predetermined research agenda; they can redirect it.

\subsection{Retention}
Scientific production generates more results than the scientific system can retain with equal prominence. Publication, peer review, citation, replication and attention therefore form another layer of selection. Peer review is itself heterogeneous. In an empirical analysis of eLife's review process, agreement between reviewers on ratings of the significance of findings and strength of support was low \parencite{mutz2025scientific}. Such disagreement is often treated simply as noise, but it also demonstrates that evaluation need not produce a single reproducible ranking. Research evaluation can itself redirect scientific activity,
as researchers and institutions adapt their behaviour to the
criteria through which performance and impact are assessed
\parencite{martin2011research}.

Once work is published, attention feeds back into future search. Highly visible papers are easier to encounter, cite, teach and build upon. Models of scientific competition show how selection for publication success can favour research practices that increase apparent productivity even when they reduce reliability \parencite{smaldino2016natural}. Models of scientific incentives also show how rewards for novel positive results can favour underpowered studies and erroneous conclusions \parencite{higginson2016current}.

Replication and other forms of verification provide a counterweight by exposing errors that production and publication processes fail to detect. The logic of replication has also been extended beyond the biomedical and social sciences to empirical work in the humanities \parencite{peels2018possibility}. Science thus contains multiple mechanisms of selection and correction. The central question is whether these mechanisms provide sufficiently distinct judgments or repeatedly transmit the same ones.

\section{Effective diversity and cross-stage dependence}
Scientific systems can appear highly diverse while providing relatively little diversity in how possibilities are explored and evaluated. Thousands of researchers may work in hundreds of institutions and publish in many journals, yet rely on similar datasets, metrics, models, funding criteria, or signals of scientific importance. Conversely, a smaller system can preserve substantially different approaches to the same uncertain problem. We use effective diversity only as descriptive shorthand for the extent to which apparently distinct components of a scientific system provide genuinely different, non-redundant routes of exploration or evaluation. We do not propose it as a separate theoretical construct or universal index. Our principal object of analysis is \textit{cross-stage dependence}.

Consider a simple example. Two initially plausible research directions, A and B, compete for attention in a field where the available evidence does not yet clearly favour either. Direction A receives a major grant. The grant allows more researchers to work on A, producing more experiments and papers. The resulting visibility makes A easier to encounter, cite and incorporate into subsequent research. In the next funding round, its publication record and prominence are themselves taken as evidence that A is the more promising direction. Recruitment, funding, publication and attention may eventually all favour A. Yet these outcomes do not necessarily represent several independent judgments that A was scientifically superior. Some later selections partly inherit the consequences of earlier ones. A may indeed be superior, but the system has generated less independent evidence about B with which to discover whether this is true.

This is the problem we call cross-stage dependence. Cross-stage dependence occurs when selection processes operating at different stages of science are not fully independent in the grounds through which they favour scientific trajectories. A decision at one stage can influence later decisions because its consequences become inputs into subsequent judgments. Dependence can also arise when different stages use the same evaluative criteria, information sources or technological infrastructure. Put simply, the question is whether a later selector contributes a genuinely new reason for favouring a trajectory, or partly inherits the reasons, signals or consequences of an earlier selection.

Cross-stage dependence should not be inferred merely from agreement or correlation between selectors. Funding agencies, journals and scientific communities may independently favour the same trajectory because each responds to strong evidence of its scientific promise. If a genuinely superior direction independently receives funding, produces convincing results and attracts subsequent attention, correlations among these outcomes do not by themselves demonstrate dependence. The empirical target is the component of later selection attributable to an earlier selection, inherited signal or shared evaluative input beyond what would be expected from independently available evidence about the scientific trajectory itself. Because scientific quality is only partially observable, experiments, quasi-experiments, longitudinal designs and comparisons across institutional settings will provide complementary rather than definitive identification.

Cross-stage dependence is related to several established ideas but differs in its object of analysis. Work on the division of cognitive labour and epistemic landscapes asks how scientists distribute effort across competing approaches and how diversity among search strategies affects collective discovery \parencite{kitcher1990division,weisberg2009epistemic,alexander2015epistemic}. Network epistemology has shown that communication structure can affect whether scientific communities preserve alternative hypotheses long enough for informative evidence to accumulate \parencite{zollman2007communication}. Work on research independence has meanwhile examined scientific autonomy, freedom, conflicts of interest and epistemic independence across disciplinary contexts \parencite{glaser2022independence}. These approaches concern relations among epistemic agents, search strategies or conditions of scientific autonomy. Cross-stage dependence instead concerns relations among functionally different mechanisms of selection: whether entry, allocation, production and retention contribute distinct grounds for deciding which scientific trajectories continue.

The concept is also distinct from cumulative advantage and path dependence. Cumulative advantage describes how earlier success changes later opportunities, as when initial funding increases the probability of subsequent funding or career success \parencite{bol2018matthew}. Path dependence describes how previous developments can constrain the subsequent production of scientific knowledge, including through theoretical and instrumental trajectories \parencite{peacock2009path}. Both can generate cross-stage dependence, but neither requires it. A trajectory may become path-dependent because specialized instruments make alternatives increasingly costly without different selectors relying on the same grounds for judgment. Conversely, funding, publication and attention may become dependent because they use overlapping prestige signals even before durable historical lock-in emerges. Cross-stage dependence therefore identifies a different property: the extent to which selections occurring at functionally different stages provide independent grounds for determining which scientific possibilities survive.
Existing evidence establishes several links through which such dependence could arise, but does not establish its prevalence as a system-level phenomenon. Early funding success can generate substantial advantages in subsequent funding among researchers with near-identical initial evaluations \parencite{bol2018matthew}. Yet propagation is not inevitable: evidence from NIH funding indicates that receiving an R01 grant produced only a modest increase in subsequent publication output, consistent with unsuccessful applicants partly substituting toward other funding sources \parencite{jacob2011impact}. Experimental evidence also shows that identical scientific work can receive substantially different evaluations depending on the prominence attributed to its author \parencite{huber2022nobel}. Random assignment of grant proposals to evaluators further shows that intellectual distance between evaluator and proposal affects evaluation, while particularly novel proposals are associated with lower scores \parencite{boudreau2016looking}. Recent systems-level synthesis likewise describes interacting pressures involving funding, productivity, reputation, career conditions and evaluation that can favour safer and more incremental research strategies \parencite{haubrock2026contemporary}. Together, these studies document mechanisms capable of transmitting or counteracting earlier selections. They do not demonstrate that contemporary science as a whole suffers from excessive cross-stage dependence.

Cross-stage dependence is not intrinsically undesirable. Scientific judgments should share reliable evidence, and once evidence strongly differentiates competing trajectories, convergence is efficient and epistemically desirable. The concern arises under uncertainty. Before the relative value of alternatives is clear, repeated filtering through overlapping grounds can remove plausible trajectories before sufficiently independent evidence has accumulated to evaluate them.

Cross-stage dependence can be investigated without requiring a universal index. One approach is to exploit experimental or quasi-experimental variation at an earlier selection stage and trace its consequences at later stages. Funding thresholds, lotteries among highly ranked proposals, or policy changes can help distinguish the effect of receiving support from characteristics that caused a proposal to be highly evaluated in the first place. Later funding, publication, recruitment or attention can then be examined to determine whether the earlier selection retains predictive power beyond scientific outputs generated subsequently. Experimental manipulation of inherited signals provides another approach: otherwise equivalent work can be evaluated with or without information about previous funding, institutional affiliation, author status, citations or other outcomes of earlier selection. A third approach is to compare systems that differ in whether successive selectors share evaluation criteria, datasets, metrics or technological infrastructure.

Evidence against cross-stage dependence is equally informative. If associations between stages disappear once independently available scientific evidence is taken into account, or if quasi-random variation in an earlier selection has no effect on later evaluation except through demonstrably improved scientific results, the proposed channel of dependence is not supported. Cross-stage dependence should therefore be treated as a family of empirically identifiable relations among selectors rather than assumed from observed concentration or agreement alone.

The framework leads to a conditional and falsifiable prediction. Among scientific systems facing substantial uncertainty and displaying comparable levels of diversity, greater cross-stage dependence should reduce the ability to explore alternative research directions and adapt when new evidence emerges, because repeated selection decisions may favour the same directions before sufficient evidence is available to assess their relative promise. Systems with greater dependence should, all else equal, eliminate plausible alternatives more rapidly, respond less readily to contradictory evidence or redirect effort more slowly when previously favoured trajectories fail. If greater measured cross-stage dependence is not associated with these outcomes under conditions of uncertainty, or systematically improves adaptive exploration, the proposed relationship would not be supported.

The aim of this research programme is not to demonstrate that dependence is intrinsically harmful. It is to determine when apparently plural scientific systems provide genuinely distinct exploration and evaluation, and when multiple stages repeatedly reproduce substantially the same selection.

\section{Artificial intelligence changes the search regime}
Artificial intelligence changes scientific search along at least three dimensions: capacity, direction, and dependence. First, AI can expand scientific capacity. Automation in science has a substantial history. Robot scientists have already demonstrated the possibility of automating parts of hypothesis formation, experimentation, and interpretation \parencite{king2009automation}. More recent systems combine large language models with laboratory tools to plan and execute chemical research tasks \parencite{boiko2023autonomous}. Contemporary work on agentic and automated scientific discovery increasingly aims to connect multiple stages of scientific reasoning and experimentation \parencite{gridach2025agentic,kramer2026automated}. These systems remain far from eliminating the physical, conceptual, and tacit constraints of scientific work, but even partial automation can change the relative cost of scientific activities.

Second, AI can change scientific direction because it does not accelerate every possibility equally. AI systems work better where problems can be represented computationally, data are accessible, objectives can be specified, and feedback can be generated. They can therefore alter which questions become cheap, tractable, or attractive. AI use has been associated with higher individual scientific productivity and impact while also being associated with greater collective concentration in the topics scientists pursue \parencite{hao2026artificial}. Increased capacity and increased breadth are therefore empirically separable.

Third, AI can alter dependence among stages. Imagine that researchers use closely related AI systems to identify literature, propose hypotheses, and choose analytical methods, while reviewers or evaluators use related systems to assess the resulting work. These remain formally distinct stages, but they may no longer provide equally distinct routes of reasoning. Biases, omissions, or representations shared by underlying models, training corpora, or ranking mechanisms could influence both what is proposed and what is subsequently judged promising. This would constitute a computational form of cross-stage dependence. It is a hypothesis, not an established consequence of AI adoption.

AI could instead increase effective diversity. Different models, specialized agents, simulation systems and search procedures could expose researchers to possibilities that human attention would otherwise miss. Automated systems could deliberately generate competing hypotheses, search neglected literatures, or challenge dominant assumptions. The direction of the effect is therefore an empirical question.
\textcite{messeri2024artificial} warn that AI can create scientific monocultures and illusions of understanding: researchers can become more productive while collectively narrowing the approaches through which scientific problems are understood. Cross-stage dependence extends this concern from the use of similar research tools to the architecture of scientific selection itself. The issue is whether common computational systems influence not only the generation of research but several mechanisms that decide which research persists.
As generation becomes cheaper, another scarcity becomes more visible: attention, verification, selection and understanding. Writing specifically about mathematics, \textcite{Gigli2026} imagines a future in which machine-generated results become abundant enough that the role of mathematicians shifts increasingly toward understanding, communication, curation, and identifying meaningful connections. This is a forward-looking disciplinary perspective rather than evidence that such a transition is inevitable, but it illustrates the distinction between producing candidate knowledge and determining what deserves sustained human attention.

AI can therefore expand search by making previously inaccessible directions tractable, intensify search within already favoured directions, or increase dependence if several stages come to rely on overlapping computational foundations. These effects can occur simultaneously. The relevant research programme is not to ask whether AI will make science better or worse in general, but to measure how it changes the distribution of scientific effort, the independence of selection processes and the capacity of communities to recover from error or redirect attention.

\section{Designing for an unknowable frontier}
The framework has practical implications, although it does not prescribe a single institutional design. If the value of scientific directions were known in advance, concentrating resources on the best trajectories would be straightforward. The problem is precisely that consequential choices are often made before such knowledge is available. Institutional design should therefore depend partly on the uncertainty surrounding competing directions.

A first principle is to preserve alternative routes when uncertainty is high. This does not mean funding every possibility equally or indefinitely. It means avoiding situations in which one early selection automatically becomes a reason for the same trajectory to win at subsequent stages. Recent proposals to diversify grant cycles, tolerate uncertainty and failure, reduce dependence on narrow evaluative proxies and decentralize parts of scientific decision-making are compatible with this principle \parencite{haubrock2026contemporary}. The present framework adds a criterion for evaluating such reforms: institutional plurality is useful for scientific search to the extent that it creates meaningfully different routes through which possibilities can survive, be tested and be reconsidered.

A second principle is to allow convergence as evidence strengthens. Independence is not an objective in itself. Requiring every funder, laboratory or journal to ignore reliable information from the rest of science would waste resources and impede accumulation. The objective should not be to prevent science from converging, but to ensure that convergence remains evidence-driven rather than being produced primarily by repeated transmission of earlier selections.

A third principle is to preserve reversibility. Scientific trajectories differ in the cost of reopening them after support disappears. A field that temporarily loses funding but retains trained researchers, data, instruments, and conceptual knowledge may remain recoverable; one whose expertise and infrastructure disappear may not. The threshold for abandoning a direction should therefore depend partly on the cost of reconstructing it. Under high uncertainty, preserving a modest capacity to return to an expensive-to-reconstruct trajectory may be more valuable than maintaining continuous large-scale investment in it.

A fourth principle concerns verification. Where possible, mechanisms intended to test scientific claims should not reproduce all the important dependencies involved in generating them. Independent datasets, methods, models, laboratories or theoretical approaches can be valuable because their failure modes differ. The same consideration applies to AI. If AI participates in generating a scientific claim, verification should, where feasible, include methods, data, models or human judgments that do not simply reproduce the same computational dependencies. This is not an argument for excluding AI from evaluation, but for recognizing shared infrastructure as a possible source of correlated error.

Reforms should therefore be evaluated not only by immediate productivity. A funding programme, peer-review reform, career policy, or AI infrastructure can increase output while reducing the number of viable routes through which unexpected evidence can redirect a field. Relevant questions include whether contradictory results receive attention, whether neglected approaches can obtain resources when new evidence supports them, whether independent methods detect common errors, how costly abandoned expertise is to reconstruct and how rapidly effort can be reallocated after a dominant approach fails.

These principles can be summarized as a conditional rule: when uncertainty is high, preserve partially independent selection, heterogeneous verification and reversible commitments; as evidence becomes stronger, greater convergence becomes justified. A resilient scientific system is not one that never converges. It is one that can still change direction after it has converged.

\section{Directions}
Science necessarily selects among more possible trajectories than it can pursue. We argue that the architecture connecting these selections deserves attention alongside the scale, quality, and diversity of scientific output. Many researchers, institutions, and evaluators do not necessarily provide many independent routes of scientific search if decisions at different stages repeatedly inherit the same signals, criteria, or technological dependencies.

Cross-stage dependence offers a way to study this architecture. Its importance is conditional rather than universal: shared evidence can produce justified convergence, while overlapping selection under substantial uncertainty can remove alternatives before they have been independently tested. Whether, where and under what conditions contemporary scientific systems exhibit enough of this dependence to impair adaptive exploration is an empirical question, not a premise of the framework.

Artificial intelligence makes the question more urgent because it can simultaneously expand what science can attempt and create common infrastructure across several stages of search and evaluation. The resulting challenge is not to maximize diversity or independence, but to preserve enough distinct routes of exploration and correction that science can respond when its current judgments are wrong.

Reality ultimately determines which scientific claims survive sufficiently rigorous confrontation with evidence. Institutions and technologies help determine which claims are attempted, which evidence is produced, and which alternatives remain available long enough to face that confrontation. The future of science will not be determined only by what remains unknown. It will also be determined by how we choose to search the unknown.

\printbibliography
\end{document}